\documentclass{iopart}

\usepackage{iopams}  
\usepackage{graphicx}

\newcommand{\ie} {{i.e., }}
\newcommand{\rmB} {{\rm B}}
\newcommand{\rmc} {{\rm c}}
\newcommand{\rmL} {{\rm L}}
\newcommand{\rms} {{\rm s}}
\newcommand{\rmT} {{\rm T}}
\newcommand{\vecf} {{\bf f}}
\newcommand{\vecr} {{\bf r}}
\newcommand{\vecu} {{\bf u}}
\newcommand{\vecv} {{\bf v}}

\newcommand{\xhat} {\hat{x}}
\newcommand{\yhat} {\hat{y}}
\newcommand{\zhat} {\hat{z}}

\begin{document}

\title [Correlated dynamics in concentrated quasi-2D suspensions]
{Correlated particle dynamics in concentrated quasi-two-dimensional
suspensions}

\author{H Diamant$^{1}$\footnote[1]{Author to whom correspondence 
should be addressed}, 
B Cui$^{2}$\footnote[2]{Present address: Department of Physics, 
Stanford University, Stanford, California 94305, USA},
B Lin$^2$ and S A Rice$^2$}

\address{$^1$ School of Chemistry, Raymond and Beverly Sackler Faculty
of Exact Sciences, Tel Aviv University,
Tel Aviv 69978, Israel}

\address{$^2$ Department of Chemistry, The James Franck Institute and
CARS, The University of Chicago, Chicago, Illinois 60637, USA}

\ead{hdiamant@tau.ac.il}

\begin{abstract}
We investigate theoretically and experimentally how the
hydrodynamically correlated lateral motion of particles in a
suspension confined between two surfaces is affected by the suspension
concentration. Despite the long range of the correlations (decaying as
$1/r^2$ with the inter-particle distance $r$), the concentration
effect is present only at short inter-particle distances for which the
static pair-correlation is nonuniform. This is in sharp contrast with
the effect of hydrodynamic screening in unconfined suspensions, where
increasing the concentration changes the prefactor of the
large-distance correlation.
\end{abstract}




\section{Introduction}
\label{sec_intro}

The Brownian motion of colloid particles and macromolecules is
correlated through hydrodynamic interactions, \ie flows that the
motion induces in the host liquid \cite{colloids,DoiEdwards}. In an
unconfined suspension \cite{Happel} these correlations decay with
inter-particle distance $r$ as $1/r$. They are positive, \ie particles
drag one another in the same direction. The long range of the
interaction leads to strong many-body effects, manifest in an
appreciable dependence of transport coefficients on concentration.  A
concentration effect of particular interest is hydrodynamic screening
\cite{DoiEdwards}, which sets in over distances much larger than the
inter-particle distance and renormalizes the prefactor of the $\sim
1/r$ pair interaction.

Colloids may sometimes be spatially confined by rigid boundaries, as
in porous media, biological constrictions, or microfluidic devices.
Attention has been turned recently toward the dynamics of such
confined suspensions \cite{Perkins}--\cite{JCP05} due to the emergence
of microfluidic applications and the development of new visualization
and manipulation techniques (digital video microscopy \cite{video} and
optical tweezers \cite{tweezer}), allowing to study dynamics at the
single- and few-particle level. The dynamics of confined suspensions
have been studied also by computer simulations
\cite{Nagele}--\cite{Finland} and have renewed the interest in related
hydrodynamic problems \cite{Mochon,Yale}.

\begin{figure}[tbh]
\centerline{\resizebox{0.4\textwidth}{!}
{\includegraphics{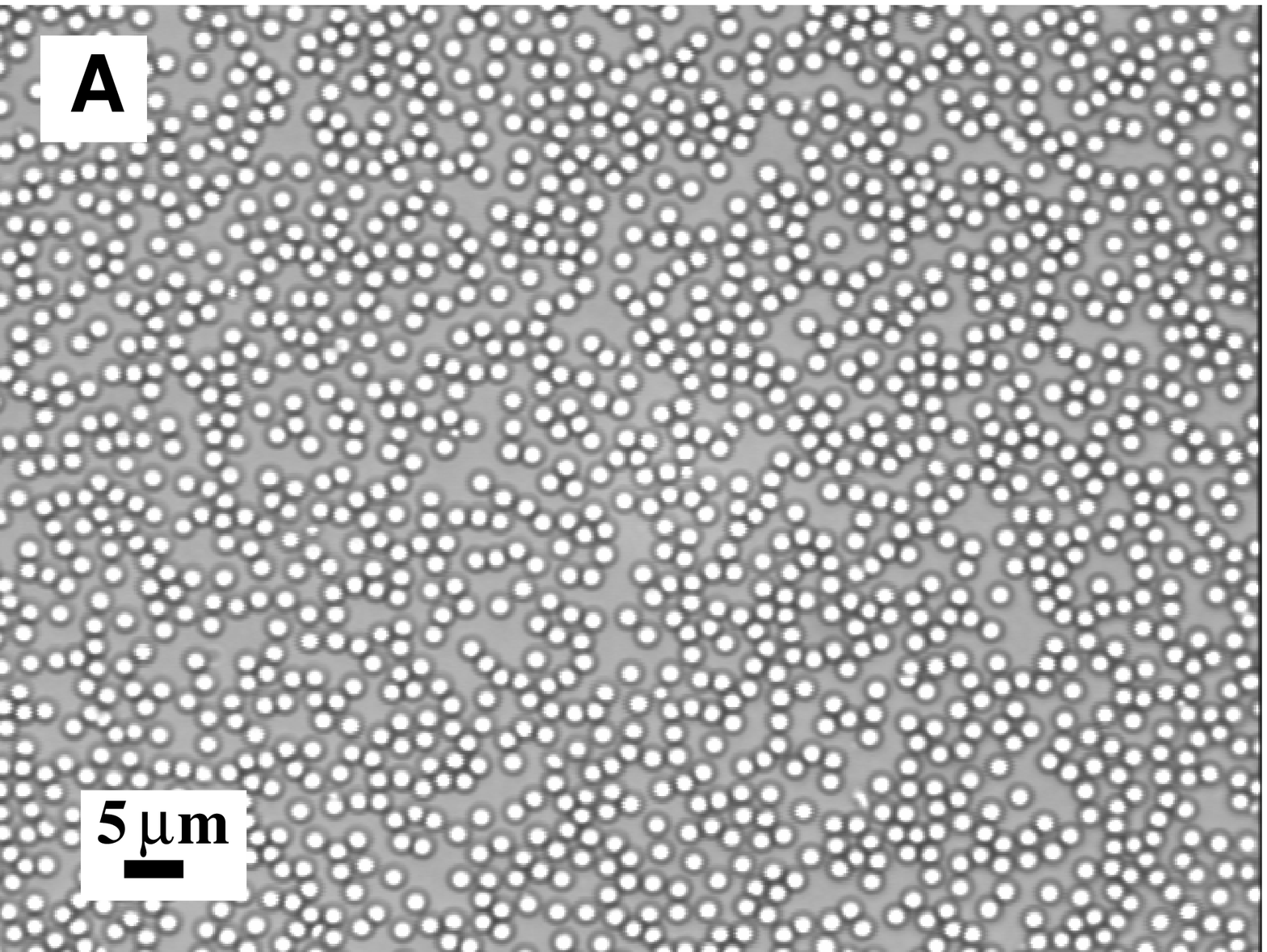}}
\hspace{0.2cm}
\resizebox{0.4\textwidth}{!}
{\includegraphics{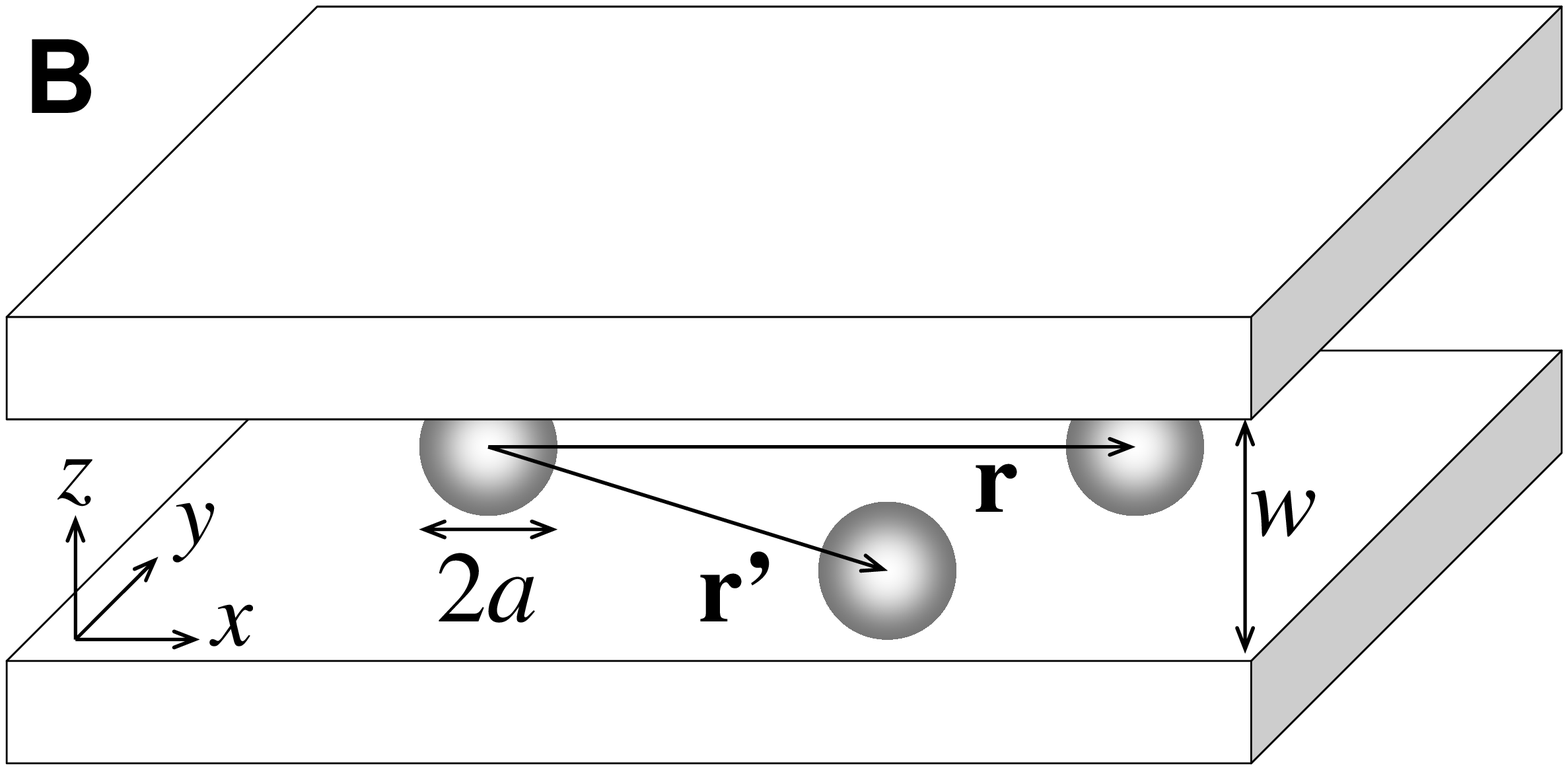}}}
\caption[]{(a) Optical microscope image 
of the experimental Q2D suspension at area fraction $\phi=0.338$.
(b) Schematic view of the system and its parameters.}
\label{fig_system}
\end{figure}

We have recently demonstrated the dramatic effect that confinement
between two flat surfaces has on the hydrodynamic pair interaction at
large inter-particle distances \cite{PRL04,JPCM05}. Such a
quasi-two-dimensional (Q2D) suspension is portrayed in figure
\ref{fig_system}.  Despite the confinement, the hydrodynamic
interaction is still long-range, decaying as $1/r^2$ rather than
$1/r$. There is ``anti-drag'' in the transverse direction, \ie the
correlation between two particles moving perpendicular to their
connecting line is negative. Arguably the most striking finding,
however, is that the concentration of the suspension has no effect on
the large-distance correlation, \ie there is no hydrodynamic
screening. These three properties are in stark contrast with their
unconfined counterparts mentioned above. 

In the current paper we review these results and then extend the
analysis to shorter inter-particle distances where the static
pair-correlation of the suspension plays a crucial role. Throughout
the paper we present the theoretical analysis along with the
corresponding experimental results.  Section \ref{sec_def} introduces
the model system and the corresponding terminology, and section
\ref{sec_exp} describes the experimental setup and methods. In section
\ref{sec_single} we briefly discuss the dynamics of a single particle
and in section \ref{sec_pair} address the hydrodynamic interaction
between two isolated particles. Section \ref{sec_three}, which
constitutes the main part of the current work, deals with the
three-body correction to the pair-interaction at finite
concentration. Finally, in section \ref{sec_dis} we conclude and
discuss the results.

\section{Model system}
\label{sec_def}

The geometry considered in this work is depicted in figure
\ref{fig_system}(b).  Identical, spherical particles of radius $a$
are suspended in a liquid of viscosity $\eta$ and temperature $T$,
confined in a slab of width $w$ between two planar solid surfaces. The
$\xhat$ and $\yhat$ axes are taken parallel, and the $\zhat$ axis
perpendicular, to the surfaces, where $z=0$ is the mid-plane. The
particles behave as hard spheres with no additional equilibrium
interaction. For simplicity we consider cases where particle motion is
restricted to two dimensions, \ie to a monolayer lying at the
mid-plane.  We use the notation $\vecr(\brho,z)$ for three-dimensional
position vectors, where $\brho(x,y)$ is a two-dimensional position
vector in the monolayer.  The area fraction occupied by particles is
denoted by $\phi$.
The Reynolds number is very low, and the hydrodynamics, therefore, are
well described by viscous Stokes flows \cite{Happel}. The confining
boundaries are impermeable and rigid, imposing no-slip boundary
conditions on the flow.

We characterize the dynamic pair correlation between two particles by
the coupling mobilities $B^\rmc_{\rm L,T}(\rho)$ as functions of the
inter-particle distance $\rho$. These are the off-diagonal terms in
the mobility tensor of a particle pair, \ie the proportionality
coefficients relating the force acting on one particle with the change
in velocity of the other.  The two independent coefficients,
$B^\rmc_{\rmL}$ and $B^\rmc_{\rmT}$, correspond, respectively, to the
coupling along and transverse to the line connecting the pair. The
dynamic correlation between two Brownian particles is similarly
characterized by two coupling diffusion coefficients,
$D^\rmc_{\rmL,\rmT}(\rho)$, which, due to the Einstein relation, are
simply related to the coupling mobility coefficients via the thermal
energy, $D^\rmc_{\rmL,\rmT}(\rho)=k_\rmB T B^\rmc_{\rmL,\rmT}(\rho)$.
[In \cite{PRL04} four coefficients were considered,
$D_{\rmL,\rmT}^\pm$, whose relation with the ones considered here is
$D^\rmc_{\rmL,\rmT}=(D_{\rmL,\rmT}^+-D_{\rmL,\rmT}^-)/2$.]

\section{Experimental setup}
\label{sec_exp}

The experimental system consists of an aqueous suspension of
monodisperse silica spheres (diameter $2a=1.58\pm 0.04$ $\mu$m,
density 2.2 g/cm$^3$, Duke Scientific), undergoing Brownian motion
while being tightly confined between two parallel glass plates in a
sealed thin cell (figure \ref{fig_system}).  The inter-plate
separation is $w=1.76\pm 0.05$ $\mu$m, \ie slightly larger than the
sphere diameter, $2a/w\simeq 0.90$.  Digital video microscopy and
subsequent data analysis are used to locate the centres of the spheres
in the field of view and then extract time-dependent two-dimensional
trajectories.  Details of the setup and measurement methods can be
found elsewhere \cite{JCP01}.  Measurements were made at four values
of area fraction, $\phi=$ 0.254, 0.338, 0.547, $0.619\pm 0.001$.
(Larger area fractions could not be checked because the suspension
began to crystallize \cite{JCP02}.)  From equilibrium studies of this
system \cite{JCP02} we infer that, for the purpose of this study, the
particles can be regarded as hard spheres.

The coupling diffusion coefficients as functions of the inter-particle
distance are directly measured from the tracked trajectories as
\begin{equation}
  D^\rmc_\rmL(\rho) = \langle x_1(t)x_2(t)\rangle_\rho/(2t),\ \ 
  D^\rmc_\rmT(\rho) = \langle y_1(t)y_2(t)\rangle_\rho/(2t), 
\label{D_exp}
\end{equation}
where $x_i(t)$ and $y_i(t)$ are the displacements of particle $i$ of
the pair during a time interval $t$ along and transverse to their
connecting line, respectively.  The average $\langle\rangle_\rho$ is
taken over all pairs whose mutual distance falls in a narrow range
($\pm 0.09$ $\mu$m) around $\rho$.

\section{Single particle}
\label{sec_single}

Consider a single particle whose centre, lying on the mid-plane
between the two confining surfaces, is defined as the origin. A force
$\vecf_1$ is applied to the particle in the $i$ direction parallel to
the surfaces ($i=x,y$). As a result, the particle moves with velocity
\begin{equation}
  u_{1i} = B_\rms f_{1i} = B_0 [(a/w)\Delta_\rms(a/w)] f_{1i},
\end{equation}
where $B_\rms$ is the self-mobility of the particle in the given
geometry and $B_0=(6\pi\eta a)^{-1}$ is its self-mobility in an
unconfined liquid.  Alternatively, we may consider a free Brownian
particle. Its mean-square displacement during a time interval $t$ will
be
\begin{equation}
  \langle\rho^2(t)\rangle = 4 D_0 [(a/w)\Delta_\rms(a/w)]t,
\end{equation}
where $D_0=k_\rmB T B_0$.  The dimensionless factor
$(a/w)\Delta_\rms(a/w)$, representing the effect of confinement,
becomes unity in the limit $a/w\ll 1$. Approximate expressions for
this factor for larger values of the confinement ratio $a/w$ were the
subject of many previous works (see \cite{Happel} and references
therein), and their validity has been confirmed in recent experiments
\cite{Lobry}--\cite{Dufresne01}.

The particle motion makes the surrounding liquid flow.  At distances
much larger than $w$ the flow velocity is given by
\cite{PRL04,JPCM05,Mochon}
\begin{eqnarray}
  v_i(\vecr) &=& (a/w)B_0 \Delta_{ij}(\vecr) f_{1j} \nonumber\\
  \Delta_{ij}(\brho,z) &=& -\lambda_0 \frac{w^2}{\rho^2}\left( \delta_{ij}
  - \frac{2\rho_i\rho_j}{\rho^2} \right) H(z/w),
\label{v_i}
\end{eqnarray}
where $i,j=x,y$, the vertical component $v_z$ is exponentially small
in $\rho/w$, and $\lambda_0$ is a dimensionless prefactor of order 1
dependent only on the confinement ratio $a/w$. The flow field
(\ref{v_i}) can be viewed as produced by an effective two-dimensional
mass dipole (source doublet), located at the particle centre and
oriented in the $j$ direction
\cite{Mochon,Ramaswamy,PRL04,JPCM05}. The flow has a perpendicular
profile $H(z/w)$ which vanishes on the two confining surfaces,
$H(-1/2)=H(1/2)=0$, and is normalized to 1 at the mid-plane, $H(0)=1$.

As argued in \cite{PRL04,JPCM05}, the far flow has the dipolar shape
(\ref{v_i}) regardless of particle size. Changing the confinement
ratio $a/w$ merely modifies the effective mass-dipole strength, \ie
the prefactor $\lambda_0(a/w)$. In the limit $a/w\rightarrow 0$ one
finds $\lambda_0=9/16$ \cite{Mochon}. As will be shown below, we find
for $a/w\simeq 0.45$ $\lambda_0\simeq 0.49$.  Since the maximum
possible confinement ratio is $a/w=1/2$, it is inferred that the
function $\lambda_0(a/w)$ actually changes very little with $a/w$.
This weak dependence on the confinement ratio can be understood in
terms of mutual compensation of two opposing effects: when $a/w$ is
larger, on one hand, the particle displaces a larger liquid volume as
it moves, yet, on the other hand, its self-mobility decreases [\ie
$(a/w)\Delta_\rms(a/w)$ gets smaller].

\section{Pair interaction}
\label{sec_pair}

Let us introduce a second particle whose centre lies at
$\vecr=(\brho,0)$. The particle is torque-free. It is also force-free
in the $xy$ plane. Being confined to the mid-plane, it evidently
cannot be assumed force-free in the $z$ direction. However, the flow
(\ref{v_i}) does not exert any force in that direction on a
mid-plane-placed particle.  Hence, as long as we restrict the
discussion to such flows, the particle can be regarded as
force-free. It is thus entrained by the flow with velocity
\cite{Happel}
\begin{equation}
  u_{2i} = \frac{1}{4\pi a^2}\int_A\rmd\vecr'' v_i(\vecr'')
  = \left(1 - \frac{4a^2}{3w^2}\right) v_i(\brho,0),
\label{Faxen1}
\end{equation}
where the integration is over the particle surface $A$. In obtaining
the last equality we have assumed a parabolic (Poiseuille) vertical
profile, $H(z/w)=1-4z^2/w^2$ \cite{Mochon}. Equation
(\ref{Faxen1}) is, in fact, a manifestation of Faxen's first law
\cite{Happel} as applied to the Q2D geometry. For a force-free particle, 
Faxen's law yields $\vecu=\vecv+(a^2/6)\nabla^2\vecv$. The dipolar
flow (\ref{v_i}) has $(\partial_{xx}+\partial_{yy})\vecv=0$ and
$\partial_{zz}|_{z=0}\vecv=[w^{-2}H''(0)/H(0)]\vecv$, from which
equation (\ref{Faxen1}) readily follows.

Substituting equation (\ref{v_i}) in equation (\ref{Faxen1}), we
obtain a relation between the force $\vecf_1$ exerted on the first
particle and the resulting velocity change $\vecu_2$ of the second
particle,
\begin{equation}
  u_{2i} = [1-4a^2/(3w^2)](a/w)B_0 \Delta_{ij}(\brho,0) f_{1j},
\label{u2}
\end{equation}
thus identifying the off-diagonal coupling terms of the pair-mobility
tensor as $B^\rmc_{ij}(\brho)=[1-4a^2/(3w^2)](a/w)B_0
\Delta_{ij}(\brho,0)$.
To get the coupling mobility along the line connecting the pair, we consider
the relation between $f_{1x}$ and $u_{2x}$ for $\brho=\rho\xhat$.
Using equation (\ref{v_i}) we get
\begin{eqnarray}
  B^\rmc_\rmL(\rho)&=&[1-4a^2/(3w^2)](a/w)B_0 \Delta_{xx}(\rho\xhat,0)
  = (a/w)B_0 \Delta_\rmL(\rho) \nonumber\\
  \Delta_\rmL(\rho)&=&\lambda w^2/\rho^2.
\label{Delta_L}
\end{eqnarray}
Similarly, for $\brho=\rho\yhat$ the relation between $f_{1x}$ and
$u_{2x}$ yields the coupling mobility transverse to the connecting
line,
\begin{eqnarray}
  B^\rmc_\rmT(\rho)&=&[1-4a^2/(3w^2)](a/w)B_0 \Delta_{xx}(\rho\yhat,0)
  = (a/w)B_0 \Delta_\rmT(\rho) \nonumber\\
  \Delta_\rmT(\rho)&=&-\lambda w^2/\rho^2.
\label{Delta_T}
\end{eqnarray}
In equations (\ref{Delta_L}) and (\ref{Delta_T}) we have defined
\begin{equation}
  \lambda(a/w) = [1-4a^2/(3w^2)]\lambda_0(a/w).
\label{lambda}
\end{equation}
This is a refinement of our previous analysis \cite{PRL04,JPCM05}. In
a Q2D geometry the so-called stokeslet approximation, equating the
coupling mobility with the flow velocity per unit force, strictly
holds only when $a$ is much smaller than {\it both} the inter-particle
distance $\rho$ and the slab width $w$, whereupon
$\lambda\simeq\lambda_0\simeq 9/16$. If $a$ is much smaller than
$\rho$ but comparable to $w$ then, no matter how large the
inter-particle distance may be, we have $\lambda<\lambda_0$. This is
due to the second term in Faxen's law, $\sim
a^2\nabla^2\vecv$. In an unconfined liquid it contributes a negligible
correction to the stokeslet limit, $O(a^2/r^2)$, whereas in Q2D
the newly introduced length $w$ leads to a significant
correction of $O(a^2/w^2)$.

As found in equations (\ref{Delta_L}) and (\ref{Delta_T}), the pair
hydrodynamic interaction in Q2D is very different from its unconfined
counterpart. The decay with distance is faster, $\sim 1/r^2$ instead
of $\sim 1/r$, yet the interaction is still long-range
\cite{Ajdari,Ramaswamy}. (Its decay, in fact, is {\it slower} than
near a single surface, where the interaction falls off as $1/r^3$
\cite{Perkins,Dufresne00}.) The transverse coupling is negative, \ie
particles exert ``anti-drag'' on one another as they move
perpendicular to their connecting line. (In the unconfined case one
has $\Delta_\rmT=\Delta_\rmL/2$, both coefficients being positive.)
These properties are direct consequences of the far flow field
(\ref{v_i}). The $1/\rho^2$ decay is that of the flow due to a 2D mass
dipole. The negative transverse coupling is a result of
circulation flows in the dipolar field \cite{PRL04}.

\begin{figure}[t]
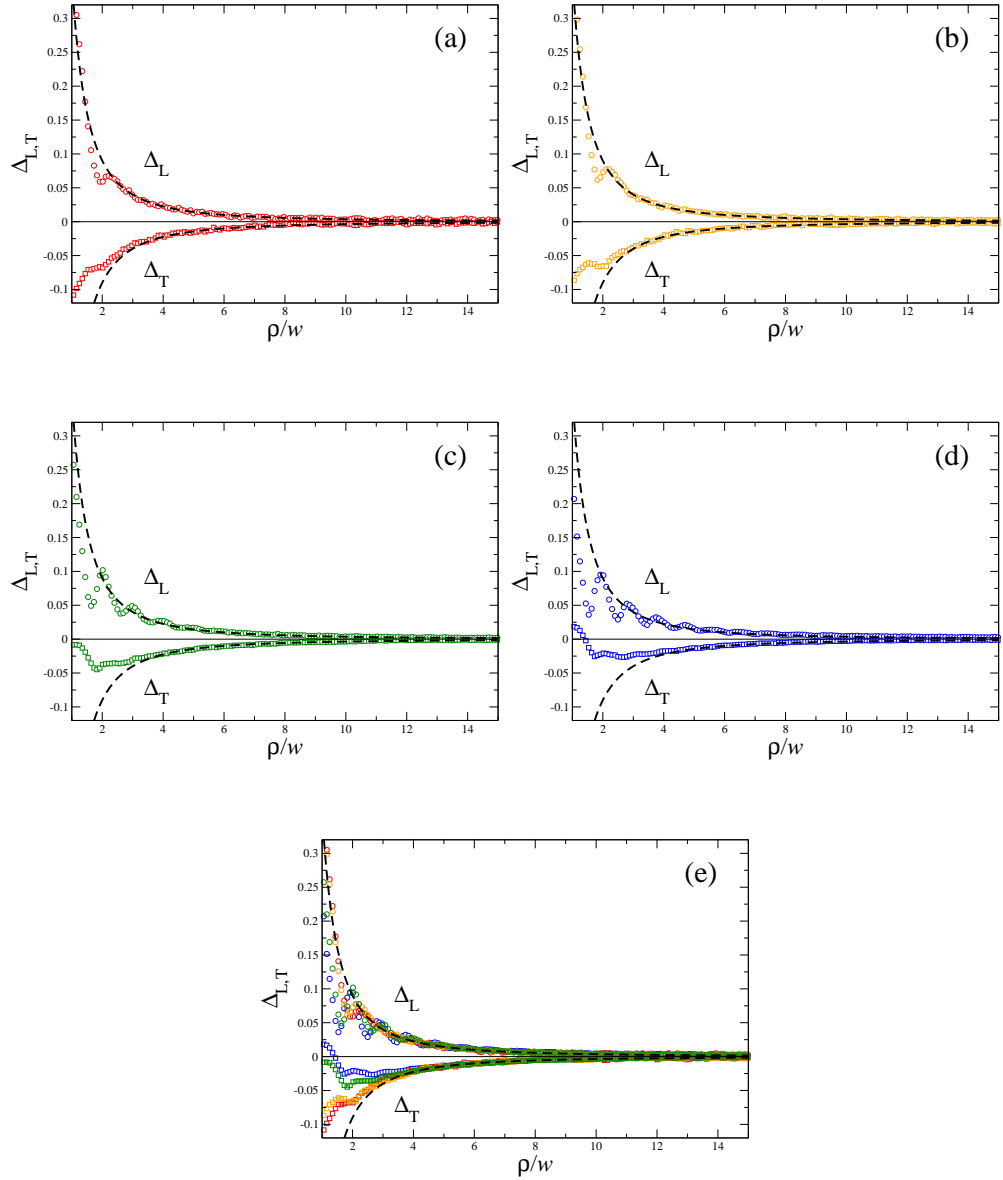

\vspace{0.7cm}
\centerline{\resizebox{0.5\textwidth}{!}
{\includegraphics{fig2a.eps}}
\resizebox{0.5\textwidth}{!}
{\includegraphics{fig2b.eps}}}
\vspace{1cm}
\centerline{\resizebox{0.5\textwidth}{!}
{\includegraphics{fig2c.eps}}
\resizebox{0.5\textwidth}{!}
{\includegraphics{fig2d.eps}}}
\vspace{1cm}
\centerline{\resizebox{0.5\textwidth}{!}
{\includegraphics{fig2e.eps}}}
\caption[]{Longitudinal ($\Delta_\rmL$, circles) and transverse
($\Delta_\rmT$, squares) coupling diffusion coefficients as a function
of inter-particle distance $\rho$. The coefficients are scaled by
$D_0a/w$ and the distance by $w$. Area fractions are $\phi=0.254$ (a,
red), 0.338 (b, orange), 0.547 (c, green), and 0.619 (d, blue).  All
data are redrawn in (e), demonstrating a concentration-independent
collapse of the large-distance measurements onto the same two curves.
Dashed lines are a fit to $\pm\lambda/(\rho/w)^2$ with the same value
of $\lambda=0.36$ for all panels.}
\label{fig_pair}
\end{figure}

For a pair of force-free Brownian particles, the hydrodynamic coupling
derived above implies that their motions be correlated according to
equation (\ref{D_exp}), with $D^\rmc_{\rmL,\rmT}(\rho)=(a/w)D_0
\Delta_{\rmL,\rmT}(\rho)$. Thus, the dimensionless couplings
$\Delta_{\rmL,\rmT}(\rho)$ can be directly measured from the
statistics of particle trajectories.  These measurements for different
area fractions $\phi$ are presented in figure \ref{fig_pair}. The
large-distance behaviour for all $\phi$ values fits well the
$\pm\lambda w^2/\rho^2$ dependence of equations (\ref{Delta_L}) and
(\ref{Delta_T}) with $\lambda\simeq 0.36$.  The negative sign of
$\Delta_\rmT$ confirms the predicted ``anti-drag'' between particles
located transverse to the direction of motion. Substituting
$\lambda\simeq 0.36$ and $a/w\simeq 0.45$ in equation (\ref{lambda})
we find $\lambda_0(0.45)\simeq 0.49$, which is only slightly smaller
than the value for a vanishing confinement ratio,
$\lambda_0(0)=9/16\simeq 0.56$.

\section{Concentration effect}
\label{sec_three}

As the area fraction $\phi$ is increased, the pair hydrodynamic
interaction should become affected by the presence of other
particles. In unconfined suspensions hydrodynamic screening sets in at
distances much larger than the typical inter-particle distance and
renormalizes the interaction by a concentration-dependent prefactor
\cite{DoiEdwards}.  Similarly, one expects the pair interaction in Q2D to
have the form of equations (\ref{Delta_L}) and (\ref{Delta_T}) yet
with a modified, $\phi$-dependent prefactor. We find, however, that
this is not the case, as is clearly demonstrated in figure
\ref{fig_pair}(e).

Consider again particle 1, located at the origin and exerting a force
$\vecf_1$ parallel to the confining surfaces.  This creates a far flow
velocity $\vecv$ according to equation (\ref{v_i}) which, in the
absence of any other particle, entrains particle 2 at
$\vecr=(\brho,0)$ with velocity $\vecu_2\sim\vecv(\vecr)$ given by
equation (\ref{u2}).  We now introduce particle 3 at
$\vecr'=(\brho',0)$. It obstructs the flow and thus modifies
$\vecu_2$.

The particle being force-free, the lowest force-distribution moment it
can exert is a force dipole,
\begin{equation}
  S_{ij}(\vecr') = \lambda_s (a/wB_0)^{-1} \frac{1}{4\pi a^2} \int_A d\vecr''
  (\vecr''-\vecr')_i v_j(\vecr''),
\label{Sij}
\end{equation}
where the integration is over the surface of particle 3, and
$\lambda_s$ is a dimensionless factor of order 1. The force dipole,
located at $\vecr'$, changes the flow velocity at $\vecr$ by
\begin{equation}
  \delta v^{s}_i(\vecr,\vecr') = S_{jk}(\vecr') (a/wB_0) \partial_j
  \Delta_{ki}(\vecr-\vecr').
\end{equation}
This holds regardless of confinement. In an unconfined system the far
flows decay as $1/r$, leading to $S\sim (r')^{-2}$ and $\delta
v^{s}\sim (r')^{-2}|\vecr-\vecr'|^{-2}$. Then, upon integration over
the third-particle position $\vecr'$, one gets a correction $\sim
1/r$, which renormalizes the coefficient of the bare $1/r$
interaction. In Q2D, however, the situation is different. First, we
realize that, due to the vanishing of $v_z$ and the symmetry of the
mid-plane, only the terms $i,j=x,y$ of $S_{ij}$ in equation
(\ref{Sij}) are non-zero. Then, since the far flows decay as
$1/\rho^2$, one gets $\delta v^{s}\sim
(\rho')^{-3}|\brho-\brho'|^{-3}$ which, upon integration over
$\brho'$, yields a correction $\sim 1/\rho^4$. At large distances this
is much smaller than the bare pair interaction ($\sim 1/\rho^2$) and
will be neglected.

Since the force dipole created by particle 3 has a negligible
far-field effect in Q2D, we proceed to the third moment of the force
distribution,
\begin{equation}
  T_{ijk}(\vecr') = \lambda_t (a/wB_0)^{-1} \frac{1}{4\pi a^2} \int_A d\vecr''
  (\vecr''-\vecr')_i (\vecr''-\vecr')_j v_k(\vecr''),
\label{Tijk}
\end{equation}
where $\lambda_t$ is another dimensionless prefactor depending only on
the confinement ratio $a/w$. In principle, $\lambda_t$ (and
$\lambda_s$) could be calculated using the {\it short-range}
hydrodynamics in a Q2D geometry \cite{Mochon,Yale}, yet we shall not
pursue this technically complicated calculation here.  The vanishing
of $v_z$ and the mid-plane symmetry make all terms $T_{ijk}$, for
which one or three of the indices are $z$, vanish. In addition, the
terms with $(i,j,k)=(x,y)$ are negligible at large distances compared
to those with two $z$ indices, since they involve two additional
powers of $1/\rho'$. We are thus left with $T_{zzi}$, $i=(x,y)$,
yielding through equations (\ref{v_i}) and (\ref{Tijk})
\begin{equation}
  T_{zzi}(\brho') = \frac{1}{3} \lambda_t a^2 \left( 1 - \frac{12a^2}{5w^2}
  \right) \Delta_{ij}(\brho',0) f_{1j}.
\label{Tijk2}
\end{equation}
In the integration we have assumed again a parabolic vertical profile,
$H(z/w)=1-4z^2/w^2$ \cite{Mochon}. Note that within this assumption
the similar leading terms in $1/\rho'$ from all higher moments of the
force distribution, involving higher $z$ derivatives, vanish.  The
moment $T(\brho')$ changes the flow velocity at $\brho$ by
\begin{equation}
  \delta v^{t}_i(\vecr,\vecr') = T_{zzj}(\brho') (a/wB_0) \partial_{zz}|_{z=0}
  \Delta_{ji}(\brho-\brho',0).
\label{deltavt}
\end{equation}
Combining equations (\ref{v_i}), (\ref{Faxen1}), (\ref{Tijk2}) and
(\ref{deltavt}), we find the correction to the velocity of particle 2
due to particle 3,
\begin{eqnarray}
  \delta u_{2i}&=&(a/wB_0) C_1
  \Delta_{ij}(\brho-\brho')\Delta_{jk}(\brho')f_{1k}
\label{deltau2}
 \\
  C_1&=&-\lambda_t \frac{8a^2}{3w^2} \left( 1 - \frac{4a^2}{3w^2}
  \right) \left( 1 - \frac{12a^2}{5w^2} \right), \nonumber
\end{eqnarray}
where the reference to $z=0$ is hereafter omitted for brevity.

Equation (\ref{deltau2}) yields the correction to $\vecu_2$ given a
certain position $\brho'$ of particle 3. We now wish to average this
correction over all possible $\brho'$,
\begin{equation}
  \langle\delta\vecu_2\rangle = \int d^2\rho' p(\brho,\brho')
  \delta\vecu_2(\brho,\brho'),
\label{deltau2av}
\end{equation}
where $p(\brho,\brho')$ is the probability density of finding particle
3 at $\brho'$ given that particle 2 is at $\brho$ and particle 1 is at
the origin. Employing the superposition approximation, we take this
probability as
\begin{equation}
  p(\brho,\brho') \simeq \frac{\phi}{\pi a^2} g(\rho')g(|\brho-\brho'|)
  \simeq \frac{\phi}{\pi a^2} [1 + h(\rho') + h(|\brho-\brho'|)],
\label{p}
\end{equation}
where $g(\rho)$ is the pair correlation function of the Q2D suspension
(normalized to 1 at $\rho\rightarrow\infty$), and $h(\rho)=g(\rho)-1$.
We have assumed that the monolayer of particles is a disordered 2D
liquid, and thus the pair correlation has no angular dependence.

Using equations (\ref{deltau2})--(\ref{p}), and specializing to the
relation between $\langle\delta u_{2x}\rangle$ and $f_{1x}$, we obtain
the average corrections to the coupling mobilities,
\begin{eqnarray}
  \delta B^{\rmc}_{\rmL,\rmT}(\rho)&=&(a/w)B_0 \delta\Delta_{\rmL,\rmT}(\rho) 
 \nonumber\\
  \delta\Delta_{\rmL,\rmT}(\rho)&=&(\pi a^2)^{-1}C_1\phi \int d^2\rho'
  [1+2h(\rho')][\Delta_{xx}(\brho-\brho')\Delta_{xx}(\brho') +
 \nonumber\\
  && \ \ \ \ \ \ \ \ \ \ \ \ \ \ \ \ \ \ \ \ 
  \Delta_{xy}(\brho-\brho')\Delta_{xy}(\brho')],
\label{deltaDelta}
\end{eqnarray}
where the longitudinal and transverse couplings are obtained, as in section
\ref{sec_pair}, by taking $\brho=\rho\xhat$ and $\brho=\rho\yhat$, 
respectively.  [Note that equation (\ref{deltaDelta}), for
$h(\rho)=0$, has exactly the same form as the one used in
\cite{PRL04,JPCM05} based on a less detailed analysis.]

A brief examination of equation (\ref{deltaDelta}) raises two
expectations. First, the integrand scales as
$(\rho')^{-2}|\brho-\brho'|^{-2}$ which, upon integration over
$\brho'$, is expected to yield a correction $\sim 1/\rho^2$, \ie a
renormalization of the prefactor of the bare coupling in accord with
hydrodynamic screening. Second, since the fields $\Delta_{ij}$ decay 
slowly with distance, one expects features (oscillations) in the static
correlation function $h(\rho)$ to be smoothed by integration and thus
be manifest only weakly in the dynamic coupling. As is shown below,
due to the unique shape of the far flow in Q2D, both of these
expectations turn out to be false.

The convolution integral of equation (\ref{deltaDelta}) is worked out in
the Appendix. The result is 
\begin{equation}
  \delta\Delta_{\rmL,\rmT}(\rho) = C \phi w^2 \left[
  \frac{h(\rho)}{\rho^2} - 2\int_\rho^\infty d\xi\frac{h(\xi)}{\xi^3} +
  O(a^2/\rho^4) \right],
\label{deltaDelta2}
\end{equation}
where $C=-2(w^2/a^2)\lambda_0^2 C_1$, and the results for the
longitudinal and transverse couplings are identical.  Equation
(\ref{deltaDelta2}) is the main result of the current work. It gives
the leading three-body correction to the pair hydrodynamic interaction
in Q2D. If static pair correlations are neglected, $h(\rho)=0$, this
correction {\it vanishes}, as was presented in \cite{PRL04,JPCM05}.
Moreover, the correction for a nonzero $h(\rho)$, although finite, is
short-range: since the suspension is disordered, the static pair
correlation $h(\rho)$ decays exponentially beyond an equilibrium
correlation length. Thus, the correction in equation
(\ref{deltaDelta2}) decreases with distance much faster than the
$1/\rho^2$ decay of the bare interaction.  This remarkable
disagreement with the usual notion of hydrodynamic screening (\ie
concentration-dependent change of the prefactor at large distances),
known from unconfined systems, is fully confirmed by our experimental
measurements as is seen in figure \ref{fig_pair}(e).

The first term in equation (\ref{deltaDelta2}) represents the leading
correction to equations (\ref{Delta_L}) and (\ref{Delta_T}) as the
inter-particle distance $\rho$ decreases. It scales as
$h(\rho)/\rho^2$ and, therefore, should be in phase with the
oscillations of the static pair correlation. The second term is
smaller, of order $ah(\rho)/\rho^3$. Higher-order corrections, which
have not been treated here, come from several sources. Corrections of
$O(a^2/\rho^4)$ arise from the integration in equation
(\ref{deltaDelta}) (see Appendix), as well as the force-dipole terms
($\delta v^s$) discussed above.  Terms of $O(h^2/\rho^2)$ come from
the approximation employed in equation (\ref{p}).  Finally, as $\rho$
becomes comparable to $w$, short-distance hydrodynamics set in and the
far flow (\ref{v_i}), on which our entire analysis has been relying,
should be corrected \cite{Mochon}.  Thus, the expression formulated in
equation (\ref{deltaDelta2}) is relevant to suspensions of
sufficiently high concentration, such that the static pair correlation
$h(\rho)$ is significant at distances $\rho>w$ where the far flow
holds.

\begin{figure}[tbh]
\vspace{0.7cm}
\centerline{\resizebox{0.6\textwidth}{!}
{\includegraphics{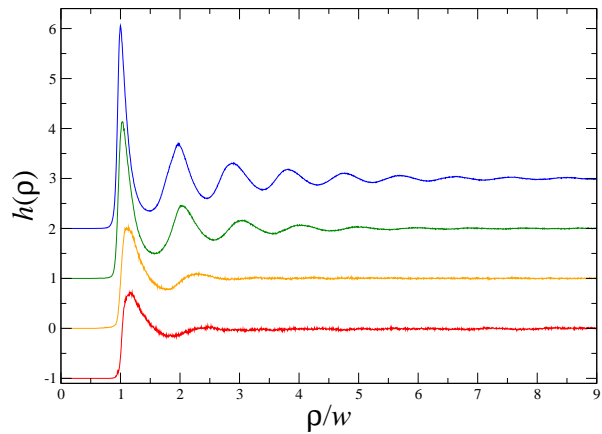}}}
\caption[]{Static pair correlation function as measured from
suspension snapshots. Plotted is $h(\rho)=g(\rho)-1$, where $g(\rho)$
is the pair correlation function normalized to 1 at
$\rho\rightarrow\infty$.  The inter-particle distance is scaled by
$w$. Area fractions are, from the bottom up, $\phi=0.254$ (red), 0.338
(orange), 0.547 (green), and 0.619 (blue).  Curves are vertically
shifted by 1 for clarity.}
\label{fig_h}
\end{figure}

To compare equation (\ref{deltaDelta2}) with experiment we need the
static pair correlation function for the various values of area
fraction. These functions are directly measurable from snapshots of
the Q2D suspensions such as the one shown in figure
\ref{fig_system}(a). (For more details, see \cite{JCP01}.)  The
results are presented in figure \ref{fig_h}.

\begin{figure}[t]
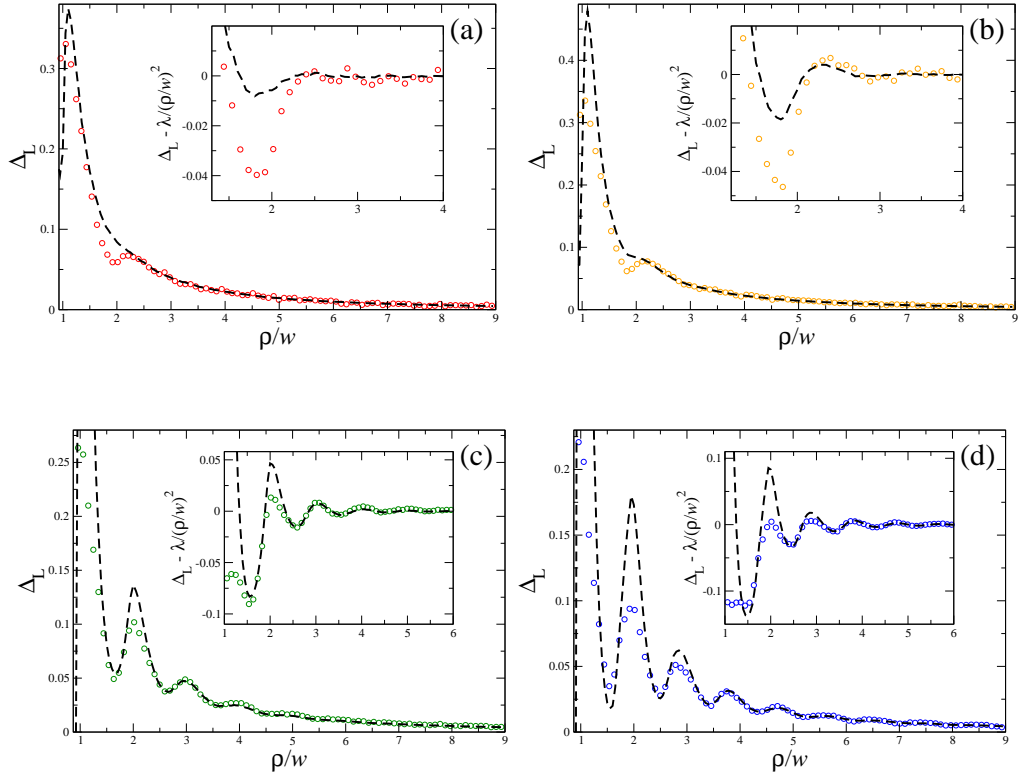

\vspace{0.7cm}
\centerline{\resizebox{0.5\textwidth}{!}
{\includegraphics{fig4a.eps}}
\hspace{0.01\textwidth}
\resizebox{0.5\textwidth}{!}
{\includegraphics{fig4b.eps}}}
\vspace{1cm}
\centerline{\resizebox{0.5\textwidth}{!}
{\includegraphics{fig4c.eps}}
\resizebox{0.5\textwidth}{!}
{\includegraphics{fig4d.eps}}}
\caption[]{Concentration effect on the longitudinal coupling diffusion
coefficient.  The diffusion coefficient is scaled by $D_0a/w$ and the
inter-particle distance by $w$. Area fractions are $\phi=0.254$ (a,
red), 0.338 (b, orange), 0.547 (c, green), and 0.619 (d, blue). Dashed
lines are a fit to equations (\ref{Delta_L}),(\ref{deltaDelta2}) with
$\lambda=0.36$ (already fitted in figure \ref{fig_pair}) and $C=0.85$
for all panels. Insets focus on the deviation of the coupling from its
long-distance behaviour. The oscillations at high concentration are in
phase with those of the measured static pair correlation (see 
figure \ref{fig_h}).}
\label{fig_wiggles}
\end{figure}

The measured $h(\rho)$ are subsequently used in equation
(\ref{deltaDelta2}) to calculate the correction to the pair
interaction. The results are shown in figure \ref{fig_wiggles}. With
the coefficient $\lambda$ of the bare interaction already found, there
is only one unknown coefficient, $C$, fitted as $C\simeq
0.85$. Working back the definitions of the various coefficients
introduced during the calculation, we find the coefficient of the
third force moment for our experimental system ($a/w\simeq 0.45$) to
be $\lambda_t\simeq 1.77$.  Comparing the fits in figure
\ref{fig_wiggles} with those in figure \ref{fig_pair}, we see that the
introduction of the concentration-dependent correction leads to a good
agreement between the theory and the measured longitudinal interaction
not only at asymptotically large distances (as was presented in
\cite{PRL04,JPCM05}) but also at intermediate distances,  
$\rho>2w$.  As anticipated, equation (\ref{deltaDelta2}) is
particularly successful for the high-concentration systems [figure
\ref{fig_wiggles}(c,d)], where oscillations related to $h(\rho)$ are
clearly observed (compare with figure \ref{fig_h}). The discrepancies
seen in figure \ref{fig_wiggles} between the theory and experiment for
$\rho\lesssim 2w$ can be attributed to several factors, as discussed
above, which have not been included in the current analysis.

\begin{figure}[tbh]
\vspace{0.7cm}
\centerline{\resizebox{0.6\textwidth}{!}
{\includegraphics{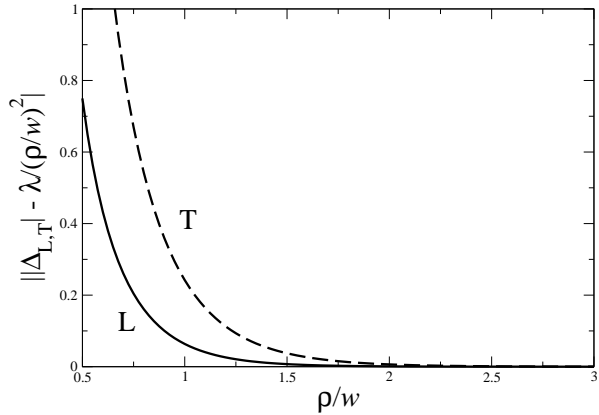}}}
\caption[]{Deviation of the pair interaction from its large-distance
behaviour in the limit of very small particles. The deviation of the
transverse interaction (T, dashed line) is larger than that of the
longitudinal one (L, solid line). The curves were calculated using the
exact solution for $a/w\ll 1$ given in \cite{Mochon}. The
large-distance behaviour in this limit is given by
$\Delta_{\rmL,\rmT}\simeq\pm\lambda/(\rho/w)^2$, with $\lambda=9/16$.}
\label{fig_stokeslet}
\end{figure}

Our calculation has yielded an identical correction to the transverse
pair interaction, $\delta\Delta_\rmT=\delta\Delta_\rmL$. This does not
agree with the measured transverse interaction, which does not exhibit
a wiggly behaviour similar to that of $\Delta_\rmL$ (see figure
\ref{fig_pair}). Thus, we cannot currently offer an accurate analysis
of the concentration effect on the transverse interaction at
intermediate and short distances. A possible reason for the difference
between the longitudinal and transverse modes may lie in the higher
sensitivity of the latter to short-range hydrodynamics.  To
demonstrate this difference we show in figure
\ref{fig_stokeslet} the deviations of the longitudinal and transverse
bare interactions from their large-distance behaviour in the limit of
very small particles ($a/w\ll 1$), as obtained from an exact solution
\cite{Mochon}. The departure of the transverse interaction from its
asymptotic behaviour is found to be more pronounced than that of the
longitudinal one.

\section{Discussion}
\label{sec_dis}

The correlated dynamics of particles in Q2D suspensions are very
different from those in unconfined systems. Confinement affects the
decay of the dynamic correlation with distance ($1/r^2$ instead of
$1/r$) as well as its sign (the correlation transverse to the line
connecting the particles becomes negative). In the current work we
have focused on the effect of concentration, \ie the presence of
additional particles, on the pair interaction. Unlike unconfined
suspensions, where the large-distance coupling changes with
concentration, in Q2D suspensions increasing the concentration has no
effect at large inter-particle distances (no hydrodynamic
screening). The qualitatively different behaviour of Q2D suspensions
has been theoretically and experimentally corroborated. It stems from
the dipolar shape of the far flow induced by particle motion. As
argued in \cite{PRL04,JPCM05,Ramaswamy}, the key property of this flow
is that it is governed by the displacement of liquid mass rather than
the diffusion of liquid momentum.

Concentration-dependent effects on the pair correlation do set in in
Q2D suspensions, yet only at intermediate and short distances. The
range of these corrections is determined by the larger of two lengths:
the range of the static pair correlation $h(\rho)$ of the suspension
and the confinement width $w$. In the former case, which is valid for
sufficiently high concentration, the spatial dependence of the dynamic
coupling reflects the features of the static pair correlation. In this
case we have been able to provide a quantitative account for the
concentration effect on the longitudinal interaction at intermediate
distances, which is in good agreement with the experimental
measurements.  In the latter case (range of $h$ shorter than $w$), the
correction depends on short-range hydrodynamics whose analysis lies
beyond the scope of the current work.

We have not treated the effect of particle motion perpendicular to the
bounding surfaces. Such fluctuations must exist in practice, yet our
results suggest that they have a minor effect at large and
intermediate distances. This is expected since the flows produced by
perpendicular fluctuations decay exponentially with distance and thus
have only a short-range effect \cite{Mochon}.

\ack

This research was supported by the Israel Science Foundation (77/03),
the National Science Foundation (CTS-021774 and CHE-9977841) and the
NSF-funded MRSEC at The University of Chicago.  H.D.\ acknowledges
additional support from the Israeli Council of Higher Education (Alon
Fellowship).

\section*{Appendix}

We need to calculate the integral appearing in equation
(\ref{deltaDelta}):
\begin{equation}
  I(\brho) = \int d^2\rho' [1+2h(\rho')]
  [\Delta_{xx}(\brho')\Delta_{xx}(\brho-\brho') +
  \Delta_{xy}(\brho')\Delta_{xy}(\brho-\brho')],
\end{equation}
where (omitting the prefactor)
\begin{equation}
  \Delta_{xx}(\brho)=(x^2-y^2)/\rho^4,\ \ \ 
  \Delta_{xy}(\brho)=2xy/\rho^4.
\label{Deltaij}
\end{equation}
Changing to polar coordinates, $\brho=(\rho,\varphi)$ and $\brho'=(\rho',\varphi')$,
we rewrite the integral as
\begin{equation}
  I(\brho) = \int d^2\rho' [1+2h(\rho')] \frac 
  {\rho^2\cos[2(\varphi'-\varphi)] + \rho'^2 - 
  2\rho\rho'\cos(\varphi'-\varphi)}
  {\rho'^2 [\rho^2 + \rho'^2 - 2\rho\rho'\cos(\varphi'-\varphi)]^2 },
\label{Ipolar}
\end{equation}
from which it is evident that $I(\brho)=I(\rho)$ is independent of
$\varphi$. Hence, we may set $\varphi=0$. This isotropy immediately
implies also that the corrections to the longitudinal and transverse
couplings, proportional to $I(\rho\xhat)$ and $I(\rho\yhat)$,
respectively, are identical.

The calculation of $I(\brho)$ via equation (\ref{Ipolar}) is a bit
tricky for reasons similar to those encountered in electrostatics. We
exclude two small areas $\sim a^2$ around the (integrable)
singularities at $\rho'=0$ and $\brho'=\brho$, and divide the
integration into three domains: (i) $\rho'\in(a,\rho-a)$,
$\varphi'\in(0,2\pi)$; (ii) $\rho'\in(\rho+a,\infty)$,
$\varphi'\in(0,2\pi)$; (iii) $\rho'\in(\rho-a,\rho+a)$,
$\varphi'\in(a/\rho,2\pi-a/\rho)$.  The contribution from the inner
domain (i) vanishes upon integration over $\varphi'$ for any $\brho$.
(Note that the static pair correlation $h(\rho)$ does not have an
angular dependence.) Integration over $\varphi'$ in the outer domain
(ii) gives $2\pi[1+2h(\rho')]/(\rho')^3$. Subsequent integration over
$\rho'$ in domain (ii) yields $2\pi\int_\rho^\infty
d\rho'[1+2h(\rho')]/(\rho')^3 + O(a/\rho^3)$. However, due to the
integrable singularity at $\brho'=\brho$, the intermediate narrow
domain (iii) has a finite contribution as well, of opposite sign:
$-\pi[1+2h(\rho)]/\rho^2 + O(a/\rho^3)$.

Thus, in the absence of static correlations, $h(\rho)=0$, the leading
terms in small $a/\rho$ from domains (ii) and (iii) cancel and
$I(\rho\gg a)=0$. This result has already been obtained in
\cite{PRL04,JPCM05}.  A more detailed inspection reveals that the
$O(a/\rho^3)$ corrections from domains (ii) and (iii) cancel as well
and, hence, $I=O(a^2/\rho^4)$. 

In the presence of static correlations, $h\neq 0$, we finally get
\begin{equation}
  I(\brho) = -2\pi \left[ \frac{h(\rho)}{\rho^2} - 2\int_\rho^\infty d\xi
  \frac{h(\xi)}{\xi^3} \right] + O(a^2/\rho^4).
\end{equation}

\section*{References}

\end{document}